\documentclass[%
 reprint,
 superscriptaddress,
 amsmath,amssymb,
]{revtex4-2}

\usepackage{graphicx}
\usepackage{dcolumn}
\usepackage{multirow} 
\usepackage{xcolor}
\usepackage[normalem]{ulem}
\usepackage{bm}
\usepackage{hyperref}

\begin{document}

\title{Bayesian phase difference estimation algorithm for direct calculation of fine structure splitting: accelerated simulation of relativistic and quantum many-body effects}

\author{Kenji Sugisaki}
\email{sugisaki@omu.ac.jp}
\affiliation{Department of Chemistry, Graduate School of Science, Osaka Metropolitan University, 3-3-138 Sugimoto, Sumiyoshi-ku, Osaka 558-8585, Japan}
\affiliation{JST PRESTO, 4-1-8 Honcho, Kawaguchi, Saitama 332-0012, Japan}
\affiliation{Centre for Quantum Engineering, Research and Education (CQuERE), TCG Centres for Research and Education in Science and Technology (TCG CREST), Sector V, Salt Lake, Kolkata 700091, India}
\author{V. S. Prasannaa}
\affiliation{Centre for Quantum Engineering, Research and Education (CQuERE), TCG Centres for Research and Education in Science and Technology (TCG CREST), Sector V, Salt Lake, Kolkata 700091, India}
\author{Satoshi Ohshima}
\affiliation{Information Technology Center, Nagoya University, Furo-cho, Chikusa-ku, Nagoya, Aichi 464-8601, Japan}
\author{Takahiro Katagiri}
\affiliation{Information Technology Center, Nagoya University, Furo-cho, Chikusa-ku, Nagoya, Aichi 464-8601, Japan}
\author{Yuji Mochizuki}
\affiliation{Department of Chemistry, Faculty of Science, Rikkyo University 3-34-1 Nishi-ikebukuro, Toshima-ku, Tokyo 171-8501, Japan}
\affiliation{Institute of Industrial Science, The University of Tokyo, 4-6-1 Komaba, Meguro-ku, Tokyo 153-8505, Japan}
\author{B. K. Sahoo}
\affiliation{Atomic, Molecular and Optical Physics Division, Physical Research Laboratory, Navrangpura, Ahmedabad 380009, India}
\author{B. P. Das}
\email{bhanu.das@tcgcrest.org}
\affiliation{Centre for Quantum Engineering, Research and Education (CQuERE), TCG Centres for Research and Education in Science and Technology (TCG CREST), Sector V, Salt Lake, Kolkata 700091, India}
\affiliation{Department of Physics, 
School of Science, Tokyo Institute of Technology, 2-12-1 Ookayama, Meguro-ku, Tokyo 152-8550, Japan} 
\date{\today}

\begin{abstract}
Despite rapid progress in the development of quantum algorithms in quantum computing as well as numerical simulation methods in classical computing for atomic and molecular applications, no systematic and comprehensive electronic structure study of atomic systems that covers almost all of the elements in the periodic table using a single quantum algorithm has been reported. In this work, we address this gap by implementing the recently-proposed quantum algorithm, the Bayesian Phase Difference Estimation (BPDE) approach, to compute accurately fine-structure splittings, which are relativistic in origin and it also depends on  quantum many-body (electron correlation) effects, of appropriately chosen states of atomic systems, including highly-charged superheavy ions. Our numerical simulations reveal that the BPDE algorithm, in the Dirac--Coulomb--Breit framework, can predict the fine-structure splitting of Boron-like ions to within 605.3 cm$^{-1}$ of root mean square deviations from the experimental ones, in the (1s, 2s, 2p, 3s, 3p) active space. We performed our simulations of relativistic and electron correlation effects on Graphics Processing Unit (GPU) by utilizing NVIDIA's cuQuantum, and observe a $\times 42.7$ speedup as compared to the CPU-only simulations in an 18-qubit active space. 
\end{abstract}

\maketitle
\section{INTRODUCTION}
Quantum computing and quantum information processing are currently among the fastest growing areas of research in modern science. In particular, recent rapid progress in the development of quantum hardware such as proof-of-principle experiments of surface code quantum error correction~\cite{Krinner2022qec, Zhao2022qec, Google2022QEC} motivates us to anticipate fault-tolerant quantum computing (FTQC) in the  future. Among the wide landscape of diverse topics in quantum computing, sophisticated ab initio electronic structure calculations of atoms and molecules on quantum computers has especially attracted much attention due to its promising real-world applications~\cite{qcqc_review_aspuru, qcqc_review_gklchan, qcqc_review_jerice}. Aiming for practical quantum computations, proper choice of a versatile quantum algorithm that can treat both light and heavy elements in the periodic table on the same footing is of crucial importance. The total energy of an atom or an atomic ion increases with atomic number. The variational quantum eigensolver (VQE)~\cite{vqe_natcomm, vqe_review_Tilly} is one of the most extensively studied algorithms for quantum chemical calculations on near-term quantum devices, but its ability to predict energies of heavier systems with a sufficiently small standard deviation is limited by a massive increase in the measurement cost. In contrast, quantum phase estimation (QPE)-based approaches~\cite{Aspuru2005, Lanyon2010, Du2010, Bauman2021} are able to compute total energies of atoms and molecules with nearly constant measurement overhead regardless of the magnitude of total energies, although development of sophisticated theoretical methods to prepare approximate wave function having sufficiently large overlap with the target electronic state is necessary~\cite{Veis2014, Sugisaki_DirCha, Sugisaki2022, Halder2022}. The quantum circuit for QPE is usually too deep to execute on a noisy intermediate-scale quantum (NISQ) device, but QPE is anticipated to be a powerful tool for electronic structure calculations of atoms and molecules in the FTQC era. A systematic study of the electronic structure of atoms and atomic systems with different atomic numbers entails accounting for the effects of special relativity, since the associated physical effects become prominent for the heavier elements. It is worth noting that even for the lighter atoms, physical phenomena that originates due to relativistic effects, such as the fine structure splitting, can be experimentally measured. However, most of the quantum simulations for quantum chemical calculations reported so far employ a non-relativistic Hamiltonian, and works in literature that take into account relativistic effects are still quite limited~\cite{Veis2012, moscovium, qed, Kasturi2022, QAE_FSS}. 

In this backdrop, we report numerical quantum simulations for the direct calculation of fine structure splitting of Boron isoeletronic sequence ($5 \leq \mathrm{Z} \leq 103$, where Z is the atomic number of the considered system) by using a Bayesian phase difference estimation (BPDE) algorithm in conjunction with the relativistic Dirac--Coulomb--Breit Hamiltonian. Fine structure splitting is, as discussed below in detail, a purely relativistic effect and is affected by electron correlation, and therefore sophisticated treatments of both relativistic and correlation effects are necessary to calculate it quantitatively. Since experimental fine-structure splitting of wide variety of isoelectronic ions have been reported~\cite{ExptlFSS}, it is a good testing ground for the sophisticated quantum chemical calculations on a quantum computer. The BPDE algorithm, which is recently proposed by one of the authors of the current work~\cite{bpde1st, bpde2nd, bpdegrad}, is a general quantum algorithm for the direct calculation of energy gaps at the full configuration interaction (FCI) level of theory and it is suitable to compute small energy differences of systems with large total energies using a quantum computer. Accelerating the computation times for quantum simulations using state-of-the-art techniques is an extremely important factor in carrying out numerical simulations of nearly a hundred systems rapidly. In this work, we report such an acceleration by utilizing Graphics Processing Unit (GPU) with NVIDIA's cuQuantum Software Developer Kit (SDK)~\cite{cuquantum}. To the best of our knowledge, this is the first comprehensive study of electronic structures of isoelectronic atomic series those covers almost all elements in the periodic table using a quantum algorithm. 

Fine structure splitting refers to the energy separation caused by the consequence of couplings between electron spin angular momentum, $S$, and the orbital angular momentum $L$, corresponding to two atomic states with different values of the total angular momentum quantum number, $J$. As an example, the electronic configuration of the ground state of Boron atom (Z = 5) in the non-relativistic scheme is $(1s)^2(2s)^2(2p)^1$. However, in the relativistic case, the degeneracy in six $p$ spin-orbitals is lifted and they split into two $p_{j=1/2}$ and four $p_{j=3/2}$ spin-orbitals. As a result, the electronic states $^2P_{1/2} = (1s_{1/2})^2(2s_{1/2})^2(2p_{1/2})^1$ and $^2P_{3/2} = (1s_{1/2})^2(2s_{1/2})^2(2p_{3/2})^1$ have different energies. The experimental value for the fine-structure splitting of B atom is  15.287 $\mathrm{cm^{-1}}$~\cite{fssBoron}, while in Boron-like Tungsten ($\mathrm{W^{69+}}$), the energy gap is $1.1802 \times 10^7 \ \mathrm{cm^{-1}}$~\cite{fssTungsten} (about six orders of magnitude larger than that of B). As pointed out above, the fine structure splitting is a purely relativistic effect and is affected by electron correlation. Accurate determination of these splittings would therefore serve as sensitive tests of relativistic many-electron theories~\cite{Das1984, Das2005, Marques2012, Dutta2012, Artemyev2013, Sahoo2019, Wan2021}. 
The extent to which the fine structure splitting is affected by electron correlation depends on the choice of system as well as the chosen states~\cite{Das1984,Das1987}. In the case of boron isoelectronic sequence, the correlation effects reduce in importance as we go from lighter to heavier ions, while for our choice of states for these systems ($^2P_{1/2}$ and $^2P_{3/2}$), pair correlation effects are the most important. In particular, the two particle-two hole excitations that result in configurations such as $(1s_{1/2})^2(2p_{1/2})^1(2p_{3/2})^2$, $(1s_{1/2})^2(2p_{1/2})^2(2p_{3/2})^1$, and $(1s_{1/2})^2(2p_{3/2})^3$ are dominant. From the view point of quantum computation, a careful choice of systems such that their fine-structure splittings span six orders of magnitude, followed by accurate determination of these quantities by a suitable quantum algorithm (BPDE in this case) would be a testament to the versatility of that algorithm.

\section{THEORY} 

\begin{figure}[t]
\includegraphics{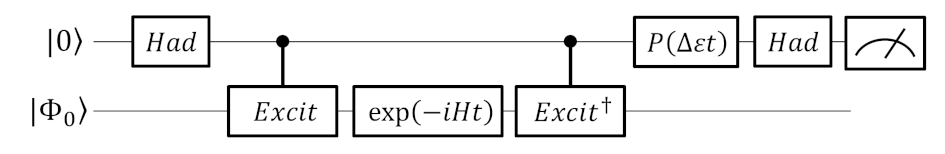}%
\caption{\label{fig:bpdecircuit} Quantum circuit for the BPDE algorithm. }
\end{figure}

A typical quantum circuit for the BPDE algorithm is illustrated in Fig.~\ref{fig:bpdecircuit}. Here, we have used the notation `\textit{Had}' for the Hadamard gate to distinguish it from the Hamiltonian, $H$.  $P(\Delta\varepsilon t)$ is a phase rotation gate defined as 
\begin{eqnarray}
P(\Delta \varepsilon t)=\left(
\begin{array}{cc}
1 & 0\\
0 & e^{i\Delta \varepsilon t}
\end{array}\right), 
\label{eq:one}
\end{eqnarray}

\noindent where $\Delta \varepsilon$ is used as the estimator of the energy gap. $|\Phi_0\rangle$ is an approximated wave function for the ground state. An approximated excited state wave function $|\Phi_1\rangle$ is generated by applying an $Excit$ gate to $|\Phi_0\rangle$. These approximated wave functions can be expanded in the basis of eigenfunctions $\{|\Psi\rangle\}$ as 
\begin{eqnarray}
|\Phi_0\rangle = \sum_j c_j |\Psi_j\rangle
\label{eq:two}
\end{eqnarray}
and 
\begin{eqnarray}
|\Phi_1\rangle = \sum_k d_k |\Psi_k\rangle. 
\label{eq:three}
\end{eqnarray} 

Using Eqs~(\ref{eq:two}) and (\ref{eq:three}), the probability of measuring the $|0\rangle$ state, $Prob(0)$, in the quantum circuit depicted in Fig 1 is calculated as 
\begin{eqnarray}
Prob(0) = \frac{1}{2}\left[1+\sum_{j,k}|c_j|^2 |d_k|^2 \mathrm{cos}\{(\Delta E_{jk} - \Delta\varepsilon)t\}\right]. 
\label{eq:four}
\end{eqnarray}  

From Eq~(\ref{eq:four}), if the approximated wave functions have sufficiently large overlap with the eigenfunction of corresponding target states, $Prob(0)$ becomes maximum around the point where $\Delta \varepsilon$ equals the energy  difference between the two targeted states. Thus, we can calculate the energy gap by finding that value of $\Delta \varepsilon$ that gives maximum $Prob(0)$. In the BPDE algorithm, $\Delta \varepsilon$ is optimized by means of Bayesian inference in the following procedure. (I) Define a prior distribution $Pr(\Delta \varepsilon)$ by a Gaussian function, in which the mean, $\mu$, corresponds to an initial estimate of the energy gap with a standard deviation $\sigma$. Note that $\sigma$ determines the energy range of the Bayesian search, and it should be large enough so that true $\Delta E$ locates between $(\mu - \sigma)$ and $(\mu + \sigma)$. (II) Repeatedly execute the quantum circuit in Fig 1 with a fixed evolution time $t = 1.8/\sigma$ and different $\Delta \varepsilon$ in the range between $(\mu - \sigma)$ and $(\mu + \sigma)$ and generate the $\Delta \varepsilon$ vs. $Prob(0)$ plot. Then, the plot is fitted by a Gaussian function and is used as a likelihood function $Pr(0|\Delta \varepsilon;t)$. (III) Calculate a posterior distribution $Pr(\Delta \varepsilon|0;t)$ using the equation 

\begin{eqnarray}
Pr(\Delta \varepsilon|0;t) = \frac{Pr(0|\Delta \varepsilon;t)Pr(\Delta \varepsilon)}{\int Pr(0|\Delta \varepsilon;t)Pr(\Delta \varepsilon)d(\Delta \varepsilon)}. 
\label{eq:five}
\end{eqnarray} 

Since both $Pr(0|\Delta \varepsilon; t)$ and $Pr(\Delta \varepsilon)$ are given as Gaussian functions, we can easily compute the posterior distribution. (IV) If the standard deviation of $Pr(\Delta \varepsilon|0;t)$ is smaller than the convergence threshold $E_{\mathrm{Thre}}$, then return the mean of $Pr(\Delta \varepsilon|0;t)$ as the estimate of $\Delta E$. Otherwise return to step (II) using the posterior distribution as the prior distribution in the next iteration.

The time evolution of wave functions is implemented using conventional techniques as follows. The second-quantized electronic Hamiltonian, built out of creation and annihilation operators (denoted below by $a^\dag$ and $a$ respectively, and with their indices running over the chosen single particle basis), is given by 

\begin{eqnarray}
H = \sum_{pq} h_{pq} a_p^\dagger a_q + \frac{1}{2} \sum_{pqrs} h_{pqrs} a_p^\dagger a_q^\dagger a_s a_r
\label{eq:six}
\end{eqnarray}

\noindent 
and is transformed to a qubit Hamiltonian, which then takes the form 

\begin{eqnarray}
H = \sum_j w_j (\sigma_{N-1}\otimes\sigma_{N-2}\otimes\cdots\sigma_0), \sigma\in\{I,X,Y,Z\}, 
\label{eq:seven}
\end{eqnarray}

\noindent 
using the Jordan--Wigner transformation~\cite{jwt}. In the equations (\ref{eq:six}) and (\ref{eq:seven}), $h_{pq}$ and $h_{pqrs}$ refer to the one- and two- electron integrals, while in the qubit Hamiltonian, $w_j$ refers to the pre-factors for each of the terms in the transformed Hamiltonian, with each $w_j$ being a product of either a one- or a two- electron integral and a multiplicative factor resulting from the transformation itself. Subsequently, the quantum circuit corresponding to the time evolution operators is constructed~\cite{Whitfield2011} using second-order Trotter--Suzuki decomposition~\cite{Trotter1959, Suzuki1976, Hatano2005book}. 

In the present study, we tested two different Hamiltonians: $H_\mathrm{DC}$ and $H_{\mathrm{DC+B}}$. $H_\mathrm{DC}$ is the Dirac--Coulomb Hamiltonian, defined by the equation (in atomic units)

\begin{eqnarray}
H_\mathrm{DC} = \sum_i [c\boldsymbol{\alpha}_i \cdot \mathbf{p}_i + (\beta_i - 1)c^2 + V_n(r_i)] + \sum_{j>i} \frac{1}{r_{ij}}, 
\label{eq:eight}
\end{eqnarray}

\noindent 
while $H_{\mathrm{DC+B}}$ contains an additional term, $V_\mathrm{B}$, given by 

\begin{eqnarray}
V_\mathrm{B} = -\sum_{j>i} \left\{\frac{\boldsymbol{\alpha}_i\cdot \boldsymbol{\alpha}_j}{2r_{ij}} 
+ \frac{(\boldsymbol{\alpha}_i\cdot\boldsymbol{\mathrm{r}}_{ij})(\boldsymbol{\alpha}_j\cdot\boldsymbol{\mathrm{r}}_{ij})}{2r_{ij}^3}\right\}
\label{eq:nine}
\end{eqnarray}

\noindent 
to take into account the full Breit interaction in addition to $H_\mathrm{DC}$. In the above equations, $c$ is the speed of light, $\boldsymbol{\alpha}$ and $\beta$ are the $4 \times 4$ Dirac matrices, $\boldsymbol{\mathrm{p}}_i$ is the momentum operator associated with the $i^{th}$ electron, $V_n(r_i)$ is the electron--nucleus potential, and $1/r_{ij}$ is the electron--electron Coulomb interaction term. A recent theoretical study by Wan and coworkers showed that relative contributions of Breit interactions to the fine-structure splitting of B isoelectronic sequence are significant for light elements, and is as large as 14.55\% for neutral Boron~\cite{Wan2021}.

For the accurate computation of fine-structure splitting, it is important to treat the $^2P_{1/2}$ ground state and the $^2P_{3/2}$ excited state on an equal footing~\cite{Das1984}. In this work, the one- and two-electron integrals were computed at the Dirac--Fock level for the state with one electron removed from the corresponding target systems (B$^+$ and its isoelectronic sequence) using our own in-house code. Gaussian-type universal basis ($\alpha_0 = 0.01$ and $\beta = 1.80$) with 40 $s_{1/2}$, 39 $p_{1/2}$, and 39 $p_{3/2}$ orbitals is used for the Dirac--Fock calculations. The large and the small components of Dirac--Fock orbitals are kinetically balanced~\cite{dyall1984}. In the BPDE simulations, we tested two types of active spaces; (1s, 2s, 2p) with 10 qubits and (1s, 2s, 2p, 3s, 3p) with 18 qubits, without adopting any qubit tapering techniques. Note that two-qubit tapering based on electron number conservation rule in the parity basis and the symmetry-conserving Bravyi--Kitaev transformation\cite{qubit_tapering} assumes non-relativistic calculations in which the electron spin quantum number is a good quantum number. Applying these techniques to relativistic calculations is not straightforward. Throughout this paper, when we use, for example, the notation (18q, $H_{\mathrm{DC+B}}$),  it specifies the active space and Hamiltonian being used. 

We now comment on the effect of Trotter error on our results. It is known that Trotter decomposition error depends on the maximum atomic charge of a system~\cite{Babbush2015}, and therefore finer Trotter decomposition should be employed for heavier elements. In the present study, we set the time length of a single Trotter step as $t/M = \max[0.2, |h_{00}|]$, where $M$ is the number of Trotter slices and $h_{00}$ is the one-electron integral corresponding to electron--nuclear attraction of the $1s_{1/2}$ electrons. We used  single Slater determinant wave functions for the approximated wave functions of the $^2P_{1/2}$ and the $^2P_{3/2}$ states: $|\Phi_0\rangle = |(1s_{1/2})^2(2s_{1/2})^2(2p_{1/2})^1\rangle$ and $|\Phi_1\rangle = |(1s_{1/2})^2(2s_{1/2})^2(2p_{3/2})^1\rangle$, respectively. In this case, the controlled-$Excit$ gate in Fig~\ref{fig:bpdecircuit} is realized by two CNOT gates. The difference between the energy expectation values of two approximated wave functions is used as the initial mean of the prior distribution: $\mu_{ini} = \Delta E_{\mathrm{Ref}} = \langle\Phi_1|H|\Phi_1\rangle - \langle\Phi_0|H|\Phi_0\rangle$, and the initial standard deviation of the prior distribution is set as $\sigma_{ini} = \max [0.1, 10|\mu_{ini}|]$ in units of Hartree. In step (II) of the BPDE algorithm, we draw 21 samples in the range of $(\mu - \sigma)$ to $(\mu + \sigma)$ with a constant interval, and execute the quantum circuit 5,000 times for each sample to construct the likelihood function. The convergence threshold for Bayesian optimization in the step (IV) is set to be inversely proportional to the time of a single Trotter step, $E_{\mathrm{Thre}} = 0.001M/t$ Hartree. These computational conditions were selected to calculate the fine-structure splitting of Boron isoelectronic sequence with similar computational costs regardless of the atomic number Z. Using these conditions, the number of Trotter slices, $M$, in the final iteration is about 1,000 for all atoms being studied.  

The numerical simulation program for the BPDE algorithm was developed using Python3 with OpenFermion\cite{OpenFermion}, Cirq\cite{cirq}, and cuQuantum\cite{cuquantum} libraries. To execute numerical simulations with GPU acceleration, we prepared the simulation environment on a Supercomputer ``Flow'' Type II subsystem in Nagoya University. The ``Flow'' Type II subsystem consists of the FUJITSU Server PRIMERGY CX2570 M5, including the Intel Xeon Gold 6230 with 20 cores $\times$ 2 sockets, and the NVIDIA Tesla V100 (Volta) $\times$ 4 GPUs, thus 33.888 TFLOPS in a node. 384 GiB (DDR4 2933 MHz) memory, and 6.4TB/node SSD in each node are available. In addition, local shared storage (BeeGFS, BeeOND, NVMesh) is also provided. The total number of nodes is 221, and hence total FLOPS is 33.888 TFLOPS $\times$ 221 nodes $=$ 7.489 PFLOPS.
The numerical simulations for 10 qubit active space were executed on Linux workstations without GPU accelerations, and those for 18 qubit active space were carried out on ``Flow'' Type II subsystem. Because the BPDE algorithm computes the likelihood function based on statistical sampling of the measurement outcome, the algorithm returns different values for every run. In this study, all the numerical simulations were carried out five times. The standard deviations of five runs for B are about 0.7--2.0 cm$^{-1}$, and those for Boron-like ions
(99 atoms in total) are on average 199.8, 163.2, 136.3, and 167.7 cm$^{-1}$ for (10q, $H_\mathrm{DC}$), (18q, $H_\mathrm{DC}$), (10q, $H_{\mathrm{DC+B}}$), and (18q, $H_{\mathrm{DC+B}}$), respectively, and they are sufficiently small compared to the calculated values of $\Delta E_{\mathrm{BPDE}}$. 

\section{RESULTS AND DISCUSSION}
\begin{figure}[b]
\includegraphics{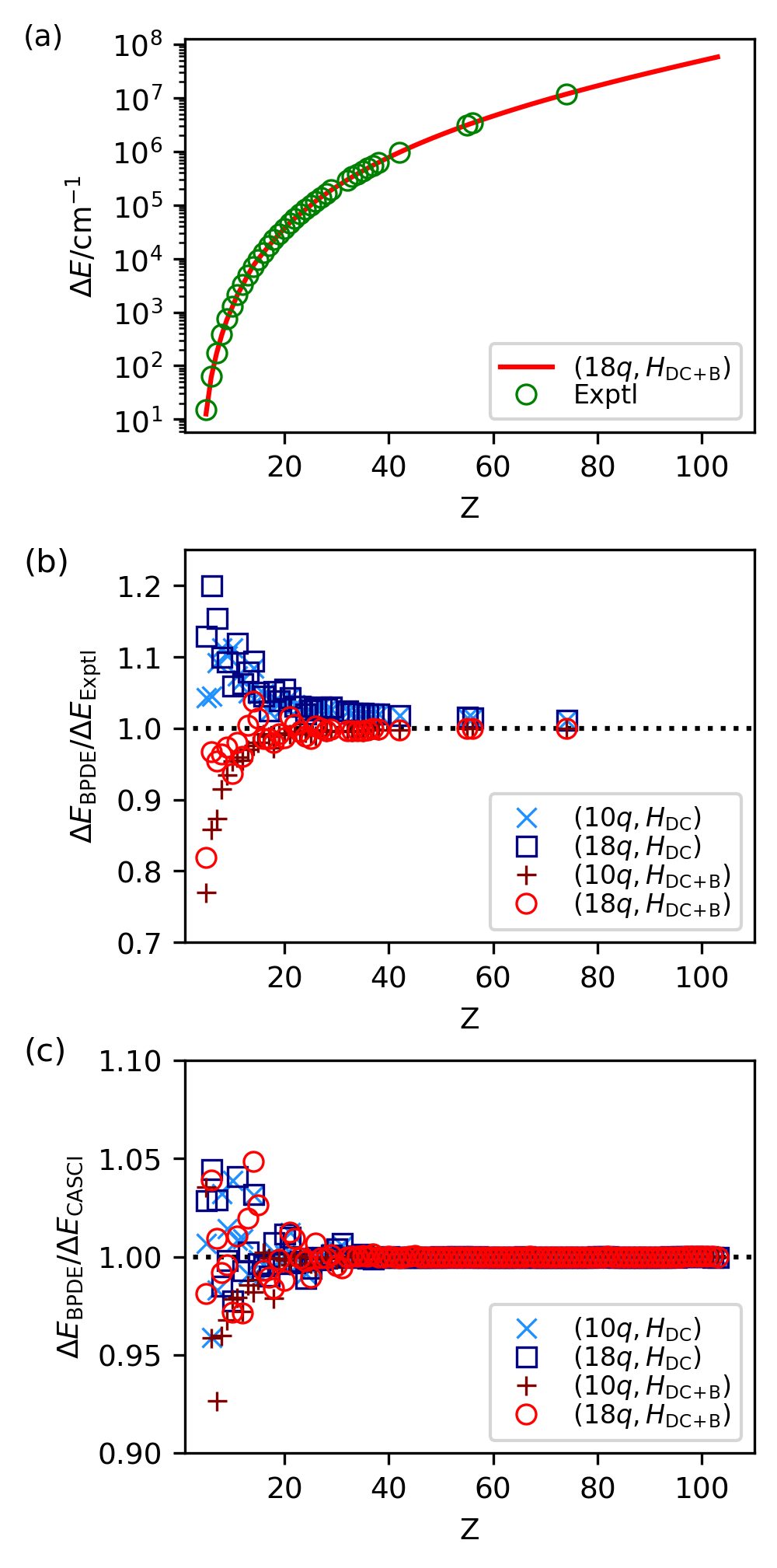}%
\caption{\label{fig:bpderesults} Results of numerical simulations of the BPDE-based fine-structure splitting calculations. (a) The fine-structure splitting values calculated by the BPDE algorithm using (18q, $H_{\mathrm{DC+B}}$) (red line) and experimental values (green circles). (b) The $\Delta E_{\mathrm{BPDE}}/\Delta E_{\mathrm{Exptl}}$ values. (c) The $\Delta E_{\mathrm{BPDE}}/\Delta E_{\mathrm{CASCI}}$ values.}
\end{figure}

The fine-structure splitting values obtained from the numerical simulations of the BPDE algorithm using (18q, $H_{\mathrm{DC+B}}$) as well as the experimental values~\cite{ExptlFSS} are plotted in Fig~\ref{fig:bpderesults}(a), and the ratios $\Delta E_{\mathrm{BPDE}}/\Delta E_{\mathrm{Exptl}}$ and $\Delta E_{\mathrm{BPDE}}/\Delta E_{\mathrm{CASCI}}$ are given in Fig~\ref{fig:bpderesults}(b) and (c), respectively. Note that CASCI is the acronym for complete active space configuration interaction, and it corresponds to the FCI treatment within the selected active orbitals. All the calculated fine-structure splitting values are provided in the Supplemental Material. For light elements, deviation of the $\Delta E_{\mathrm{BPDE}}$ values from experimental ones appears to be large, but this observation is a consequence of the fine-structure splitting being small. In fact, the fine-structure splitting of B calculated using the BPDE algorithm with (18q, $H_{\mathrm{DC+B}}$) is $\Delta E = 12.5287
\pm 0.9813\ \mathrm{cm^{-1}}$, and the absolute error with respect to the experimental value is only  about 3 cm$^{-1}$. Note that the $\Delta E$ value of B calculated at the CASCI level is 12.2715 cm$^{-1}$ and therefore departure of the $\Delta E_{\mathrm{BPDE}}$ value from the experimental one is well explained by the limited active space size. Note that all the $\Delta E_{\mathrm{BPDE}}/\Delta E_{\mathrm{CASCI}}$ values (Fig~\ref{fig:bpderesults}c) are in the range 0.92 and 1.05, thus indicating the ability of the BPDE algorithm in being able to reproduce the CASCI fine-structure splitting very accurately. We expect that inclusion of more virtual orbitals (3d, 4s, 4p, and above) can further improve the accuracy of the  fine-structure splitting. 
The agreement in the values of $\Delta E_{\mathrm{BPDE}}$ with $\Delta E_{\mathrm{Exptl}}$ in highly charged ions is also worth emphasizing. In this context, we note that for heavier ions,  relativistic effects are dominant and electron correlation effects becomes less significant. Nevertheless, the fine-structure splittings computed by using the BPDE algorithm are closer to the experimental values than those estimated from the reference wave functions $\Delta E_{\mathrm{Ref}}$. For example, the fine-structure splitting in Boron-like Tungsten (W$^{69+}$) is calculated to be $\Delta E_{\mathrm{Ref}} = 11,841,730\ \mathrm{cm^{-1}}$ and $\Delta E_{\mathrm{BPDE}} = 11,800,183\ \mathrm{cm^{-1}}$ for (18q, $H_{\mathrm{DC+B}}$), and the experimental value is $\Delta E_{\mathrm{Exptl}} = 11,802,000\ \mathrm{cm^{-1}}$. These results also exemplify the capability of the BPDE algorithm in predicting the energy gap accurately. The root mean square deviations of the $\Delta E_{\mathrm{BPDE}}$ from $\Delta E_{\mathrm{Exptl}}$ were calculated as 9966.6, 10139.7, 698.0, and 605.3 cm$^{-1}$ for (10q, $H_\mathrm{DC}$), (18q, $H_\mathrm{DC}$), (10q, $H_{\mathrm{DC+B}}$), and (18q, $H_{\mathrm{DC+B}}$), respectively. It is worth noting that the larger active space with more sophisticated Hamiltonian (18q, $H_{\mathrm{DC+B}}$) gives the best agreement with the experimental values. 

In order to check for GPU acceleration in our numerical quantum circuit simulation, we executed the BPDE calculations of Boron fine-structure splitting with 8, 10, 16, and 18 qubit active spaces in conjunction with $H_{\mathrm{DC+B}}$ on our Linux workstation (CPU: Intel Xeon-Gold 6134, GPU: None) and ``Flow'' Type II subsystem (CPU: Intel Xeon-Gold 6230, GPU: NVIDIA Tesla V100). The 8 and 16 qubit active spaces were prepared by fixing the occupation number of $1s_{1/2}$ orbitals in the 10 and 18 qubit active spaces, respectively. For 8 and 10 qubit active spaces, we also carried out the BPDE simulations on ``Flow'' Type II subsystem without using GPU. All the simulations were performed five times using single thread. 
\begin{table}
\caption{\label{tab:table1} Average time taken for BPDE quantum circuit simulation for B atom fine-structure splitting, in units of seconds. 
}
\begin{ruledtabular}
\begin{tabular}{cccc}
\multirow{2}{*}{Size of active space} & Workstation\footnote{CPU: Intel Xeon-Gold 6134, GPU: None} & \multicolumn{2}{c}{``Flow'' Type II\footnote{CPU: Intel Xeon-Gold 6230, GPU: NVIDIA Tesla V100}}\\
& w/o GPU & w/o GPU & with GPU\\
\hline
8 & 628 & 731 & 177 \\
10 & 2197 & 2267 & 588 \\
16 & 73452 & --- & 4830 \\
18 & 387328 & --- & 9081 \\
\end{tabular}
\end{ruledtabular}
\end{table}

The average simulation time of five runs are given in Table~\ref{tab:table1}. From the table, the speedup in GPU-accelerated quantum circuit simulations is significant, especially when a larger active space is employed. For smaller active spaces (8q and 10q), the speedup is about $\times4$. By contrast, for 16 and 18 qubit active spaces,  GPU-based simulations provide substantial speedups of 15.2 and 42.7 times, respectively. Note that the computation times of both CPU and GPU-based quantum circuit simulations scales exponentially with the number of qubits, but the exponent is smaller for GPU than CPU. 

\section{SUMMARY}
In summary, our numerical quantum simulations reveal that quantum computers have the potential of computing fine-structure splitting of Boron isoelectronic sequence very accurately by appropriately considering both relativistic and quantum many-body effects using the Bayesian phase difference estimation algorithm. By using the (1s, 2s, 2p, 3s, 3p) active space and the relativistic Dirac--Coulomb--Breit Hamiltonian, the fine-structure splittings in Boron isoelectronic sequence were predicted within 605.3 cm$^{-1}$ of the root mean square deviation from the experimental values. This root mean square deviation value is the smallest among the four cases involving different combinations of Hamiltonian and number of qubits that we tested. It is noteworthy that the deviations in the calculated fine structure splittings from the experimental values 
do not increase nearly as much as they do with increase in atomic number. 
Two of our important findings in this work regarding accuracy of the calculated fine structure splitting are (a) that it is crucial to include Breit interactions for accurate predictions of fine-structure splittings for the systems that we have considered, and (b) that it is necessary to include virtual orbitals to improve our results for the lighter systems (neutral Boron and the lighter Boron-like ions). Speedup in numerical simulations of quantum circuits by using GPU in conjunction with NVIDIA cuQuantum is significant, especially when large active space is employed. We observed $\times 42.7$ speedup for 18-qubit active space simulations. Such acceleration of quantum circuit simulations is important to test the ability of the quantum algorithms to handle problems of larger size, for further development of quantum algorithms for various applications. 

\begin{acknowledgments}
The authors thank Shinya Morino (NVIDIA Japan) for helpful discussions on the cuQuantum simulation environment construction on ``Flow'' Type II subsystem. This work was supported by JST PRESTO ``Quantum Software'' project (Grant No. JPMJPR1914), Japan and KAKENHI Scientific Research C (Grant No. 21K03407) from JSPS, Japan. The computation was carried out using the JHPCN Joint Research Projects (jh220010) on supercomputer ``Flow'' at Information Technology Center, Nagoya University.
\end{acknowledgments}

\bibliography{references}
\bibliographystyle{unsrt}

\newpage
\renewcommand{\thetable}{S1} 
\begin{table*}
\textbf{SUPPLEMENTAL MATERIAL}
\caption{Fine structure splitting of Boron isoelectronic sequence calculated using $H_{\mathrm{DC}}$.}
\begin{ruledtabular}
\begin{tabular}{ccccccc}
Atom & $\Delta E_{\mathrm{Ref}}$ & $\Delta E_{\mathrm{CASCI}(10q)}$ & $\Delta E_{\mathrm{BPDE}(10q)}$ & $\Delta E_{\mathrm{CASCI}(18q)}$ & $\Delta E_{\mathrm{BPDE}(18q)}$ & $\Delta E_{\mathrm{Exptl}}$\footnote{NIST Atomic spectra database levels data. \url{https://physics.nist.gov/PhysRefData/ASD/levels_form.html.} Retrieved: 2022-08-22.}   \\
\hline
B & 17.9174 & 15.8486 & 15.9481 $\pm$ 1.2618 & 16.7709 & 17.2684 $\pm$ 1.9425 & 15.287 \\
C$^+$ & 74.868 & 69.831 & 66.296 $\pm$ 1.574 & 71.989 & 76.088 $\pm$ 6.982 & 63.42 \\ 
N$^{2+}$ & 202.79 & 192.26 & 190.42 $\pm$ 0.98 & 195.97 & 201.23 $\pm$ 2.98 & 174.4 \\
O$^{3+}$ & 442.28 & 422.50 & 429.59 $\pm$ 9.91 & 428.09 & 424.54 $\pm$ 6.25 & 385.9 \\
F$^{4+}$ & 842.74 & 808.52 & 820.26 $\pm$ 2.66 & 816.38 & 813.68 $\pm$ 5.11 & 744.5 \\
Ne$^{5+}$ & 1456.56 & 1407.02 & 1455.09 $\pm$ 32.58 & 1417.54 & 1384.46 $\pm$ 4.05 & 1306.81 \\
Na$^{6+}$ & 2368.89 & 2283.50 & 2292.26 $\pm$ 37.54 & 2297.12 & 2390.01 $\pm$ 4.71 & 2134.61 \\
Mg$^{7+}$ & 3638.30 & 3512.34 & 3528.87 $\pm$ 51.17 & 3529.50 & 3508.24 $\pm$ 7.99 & 3302 \\
Al$^{8+}$ & 5356.23 & 5176.92 & 5132.37 $\pm$ 23.11 & 5198.09 & 5276.66 $\pm$ 49.49 & 4890 \\
Si$^{9+}$ & 7617.46 & 7369.72 & 7582.29 $\pm$ 281.69 & 7395.39 & 7647.06 $\pm$ 28.88 & 6990.6 \\
P$^{10+}$ & 10526.07  & 10192.46 & 10175.27 $\pm$ 5.90 & 10223.13 & 10192.56 $\pm$ 7.21 & 9699 \\
S$^{11+}$ & 14195.56 & 13756.22 & 13781.11 $\pm$ 139.22 & 13792.41 & 13734.39 $\pm$ 120.46 & 13135.3 \\
Cl$^{12+}$ & 18749.02 & 18181.61 & 17827.99 $\pm$ 32.46 & 18223.86 & 17842.14 $\pm$ 88.91 & 17410 \\
Ar$^{13+}$ & 24319.18 & 23598.96 & 23508.34 $\pm$ 260.68 & 23647.82 & 23843.43 $\pm$ 141.14 & 22656.239 \\
K$^{14+}$ & 31048.63 & 30148.47 & 30036.00 $\pm$ 103.44 & 30204.49 & 30149.39 $\pm$ 95.46 & 29017 \\
Ca$^{15+}$ & 39089.90 & 37980.39 & 37647.62 $\pm$ 301.63 & 38044.15 & 38530.65 $\pm$ 280.51 & 36520 \\
Sc$^{16+}$ & 48605.67 & 47255.28 & 47680.40 $\pm$ 877.20 & 47327.35 & 47617.66 $\pm$ 314.80 & 45637 \\
Ti$^{17+}$ & 59768.90 & 58144.16 & 58117.48 $\pm$ 145.09 & 58225.12 & 58024.71 $\pm$ 51.44 & 56240 \\
V$^{18+}$ & 72763.06 & 70828.77 & 70739.20 $\pm$ 157.39 & 70919.22 & 70751.88 $\pm$ 77.13 & 68610 \\
Cr$^{19+}$ & 87782.29 & 85501.81 & 84842.85 $\pm$ 442.65 & 85602.32 & 84460.94 $\pm$ 212.75 & 82970 \\
Mn$^{20+}$ & 105032 & 102367 & 101745 $\pm$ 220 & 102478 & 101935 $\pm$ 481 & 99360 \\
Fe$^{21+}$ & 124727 & 121640 & 121629 $\pm$ 18 & 121762 & 121769 $\pm$ 69 & 118266 \\
Co$^{22+}$ & 147097 & 143548 & 143592 $\pm$ 54 & 143682 & 143428 $\pm$ 138 & 139290 \\
Ni$^{23+}$ & 172379 & 168329 & 168364 $\pm$ 301 & 168475 & 168480 $\pm$ 159 & 163960 \\
Cu$^{24+}$ & 200825 & 196234 & 196465 $\pm$ 77 & 196394 & 197027 $\pm$ 196 & 191280 \\
Zn$^{25+}$ & 232698 & 227528 & 228371 $\pm$ 964 & 227701 & 228747 $\pm$ 310 & \\
Ga$^{26+}$ & 268274 & 262486 & 263850 $\pm$ 592 & 262673 & 264403 $\pm$ 35 & \\
Ge$^{27+}$ & 307841 & 301399 & 301484 $\pm$ 262 & 301600 & 301747 $\pm$ 223 & 294550 \\
As$^{28+}$ & 351700 & 344568 & 344740 $\pm$ 249 & 344784 & 344750 $\pm$ 96 & 337400 \\
Se$^{29+}$ & 400166 & 392311 & 392679 $\pm$ 81 & 392543 & 392775 $\pm$ 61 & 384460 \\
Br$^{30+}$ & 453569 & 444958 & 445279 $\pm$ 373 & 445206 & 445945 $\pm$ 42 & 436400 \\
Kr$^{31+}$ & 512250 & 502853 & 502510 $\pm$ 453 & 503117 & 502733 $\pm$ 668 & 492560 \\
Rb$^{32+}$ & 576568 & 566358 & 565943 $\pm$ 269 & 566639 & 566283 $\pm$ 431 & 554700 \\
Sr$^{33+}$ & 652881 & 635846 & 635884 $\pm$ 182 & 636144 & 636082 $\pm$ 204 & 623100 \\
Y$^{34+}$ & 723619 & 711707 & 711912 $\pm$ 113 & 712023 & 711874 $\pm$ 141 & \\
Zr$^{35+}$ & 807144 & 794349 & 794633 $\pm$ 35 & 794683 & 794833 $\pm$ 122 & \\
Nb$^{36+}$ & 897891 & 884195 & 883919 $\pm$ 151 & 884547 & 884216 $\pm$ 69 & \\
Mo$^{37+}$ & 996297 & 981684 & 981773 $\pm$ 145 & 982055 & 981726 $\pm$ 559 & 964360 \\
Tc$^{38+}$ & 1102818 & 1087273 & 1087296 $\pm$ 427 & 1087664 & 1087542 $\pm$ 275 & \\
Ru$^{39+}$ & 1217926 & 1201440 & 1201283 $\pm$ 284 & 1201850 & 1201776 $\pm$ 291 & \\
Rh$^{40+}$ & 1342115 & 1324677 & 1324715 $\pm$ 100 & 1325107 & 1325094 $\pm$ 122 & \\
Pd$^{41+}$ & 1475895 & 1457498 & 1457410 $\pm$ 220 & 1457949 & 1457916 $\pm$ 230 & \\
Ag$^{42+}$ & 1619798 & 1600438 & 1600596 $\pm$ 254 & 1600910 & 1600721 $\pm$ 86 & \\
Cd$^{43+}$ & 1774377 & 1754051 & 1754295 $\pm$ 91 & 1754543 & 1754580 $\pm$ 225 & \\
In$^{44+}$ & 1940207 & 1918912 & 1918639 $\pm$ 115 & 1919427 & 1919129 $\pm$ 29 & \\
Sn$^{45+}$ & 2117884 & 2095621 & 2096259 $\pm$ 90 & 2096158 & 2096422 $\pm$ 129 & \\
Sb$^{46+}$ & 2308028 & 2284799 & 2284125 $\pm$ 143 & 2285359 & 2284924 $\pm$ 176 & \\
Te$^{47+}$ & 2511283 & 2487091 & 2487548 $\pm$ 194 & 2487673 & 2487648 $\pm$ 135 & \\
I$^{48+}$ & 2728321 & 2703171 & 2703597 $\pm$ 192 & 2703777 & 2704096 $\pm$ 141 & \\
Xe$^{49+}$ & 2959835 & 2933731 & 2934128 $\pm$ 196 & 2934362 & 2934422 $\pm$ 197 & \\
Cs$^{50+}$ & 3206549 & 3179500 & 3179630 $\pm$ 128 & 3180155 & 3180328 $\pm$ 166 & 3131500 \\
Ba$^{51+}$ & 3469213 & 3441225 & 3440903 $\pm$ 52 & 3441905 & 3441680 $\pm$ 197 & 3390000 \\
La$^{52+}$ & 3748611 & 3719693 & 3719951 $\pm$ 86 & 3720399 & 3720606 $\pm$ 163 & \\
Ce$^{53+}$ & 4045553 & 4015715 & 4015654 $\pm$ 116 & 4016448 & 4016555 $\pm$ 158 & \\
Pr$^{54+}$ & 4360884 & 4330136 & 4330527 $\pm$ 193 & 4330896 & 4331133 $\pm$ 123 & \\
Nd$^{55+}$ & 4695479 & 4663830 & 4663813 $\pm$ 139 & 4664619 & 4664367 $\pm$ 140 & \\
\end{tabular}
\end{ruledtabular}
\end{table*}

\begin{table*}
\begin{ruledtabular}
\begin{tabular}{ccccccc}
Atom & $\Delta E_{\mathrm{Ref}}$ & $\Delta E_{\mathrm{CASCI}(10q)}$ & $\Delta E_{\mathrm{BPDE}(10q)}$ & $\Delta E_{\mathrm{CASCI}(18q)}$ & $\Delta E_{\mathrm{BPDE}(18q)}$ & $\Delta E_{\mathrm{Exptl}}$\footnote{NIST Atomic spectra database levels data. \url{https://physics.nist.gov/PhysRefData/ASD/levels_form.html.} Retrieved: 2022-08-22.}   \\
\hline
Pm$^{56+}$ & 5050254 & 5017718 & 5017925 $\pm$ 305 & 5018534 & 5018126 $\pm$ 137 & \\
Sm$^{57+}$ & 5426153 & 5392737 & 5393112 $\pm$ 208 & 5393583 & 5393480 $\pm$ 235 & \\
Eu$^{58+}$ & 5824171 & 5789889 & 5789980 $\pm$ 52 & 5790766 & 5790877 $\pm$ 124 & \\
Gd$^{59+}$ & 6245328 & 6210189 & 6210225 $\pm$ 405 & 6211097 & 6211030 $\pm$ 57 & \\
Tb$^{60+}$ & 6690696 & 6654722 & 6654865 $\pm$ 169 & 6655662 & 6655651 $\pm$ 360 & \\
Dy$^{61+}$ & 7161401 & 7124590 & 7126185 $\pm$ 212 & 7125563 & 7126567 $\pm$ 75 & \\
Ho$^{62+}$ & 7658589 & 7620961 & 7620883 $\pm$ 202 & 7621967 & 7621731 $\pm$ 111 & \\
Er$^{63+}$ & 8183474 & 8145040 & 8145092 $\pm$ 530 & 8146081 & 8145938 $\pm$ 238 & \\
Tm$^{64+}$ & 8737316 & 8698090 & 8697909 $\pm$ 279 & 8699168 & 8698577 $\pm$ 226 & \\
Yb$^{65+}$ & 9321420 & 9281411 & 9281615 $\pm$ 161 & 9282526 & 9282366 $\pm$ 53 & \\
Lu$^{66+}$ & 9937169 & 9896391 & 9896412 $\pm$ 258 & 9897545 & 9897147 $\pm$ 157 & \\
Hf$^{67+}$ & 10585979 & 10544442 & 10543803 $\pm$ 123 & 10545635 & 10544828 $\pm$ 179 & \\
Ta$^{68+}$ & 11269350 & 11227067 & 11226933 $\pm$ 418 & 11228302 & 11228238 $\pm$ 42 & \\
W$^{69+}$ & 11988832 & 11945815 & 11945395 $\pm$ 227 & 11947092 & 11946340 $\pm$ 65 & 11802000 \\
Re$^{70+}$ & 12746058 & 12702317 & 12702908 $\pm$ 91 & 12703638 & 12703808 $\pm$ 86 & \\
Os$^{71+}$ & 13542712 & 13498254 & 13498082 $\pm$ 240 & 13499621 & 13498828 $\pm$ 266 & \\
Ir$^{72+}$ & 14380594 & 14335433 & 14336439 $\pm$ 189 & 14336847 & 14337502 $\pm$ 63 & \\
Pt$^{73+}$ & 15261542 & 15215686 & 15215990 $\pm$ 198 & 15217149 & 15217012 $\pm$ 125 & \\
Au$^{74+}$ & 16187517 & 16140977 & 16141619 $\pm$ 228 & 16142490 & 16142679 $\pm$ 39 & \\
Hg$^{75+}$ & 17160527 & 17113306 & 17115189 $\pm$ 231 & 17114872 & 17116484 $\pm$ 347 & \\
Tl$^{76+}$ & 18182718 & 18134822 & 18136036 $\pm$ 139 & 18136442 & 18137391 $\pm$ 180 & \\
Pb$^{77+}$ & 19256340 & 19207778 & 19209871 $\pm$ 306 & 19209454 & 19212153 $\pm$ 54 & \\
Bi$^{78+}$ & 20383744 & 20334524 & 20334279 $\pm$ 175 & 20336259 & 20335669 $\pm$ 296 & \\
Po$^{79+}$ & 21567411 & 21517545 & 21517352 $\pm$ 149 & 21519341 & 21517098 $\pm$ 554 & \\
At$^{80+}$ & 22809881 & 22759368 & 22758762 $\pm$ 167 & 22761228 & 22760323 $\pm$ 135 & \\
Rn$^{81+}$ & 24113621 & 24062411 & 24061957 $\pm$ 95 & 24064337 & 24063545 $\pm$ 183 & \\
Fr$^{82+}$ & 25481984 & 25430126 & 25430910 $\pm$ 243 & 25432121 & 25432069 $\pm$ 350 & \\
Ra$^{83+}$ & 26917680 & 26865161 & 26864880 $\pm$ 96 & 26867227 & 26865772 $\pm$ 189 & \\
Ac$^{84+}$ & 28423992 & 28370819 & 28371577 $\pm$ 241 & 28372960 & 28372993 $\pm$ 133 & \\
Th$^{85+}$ & 30004040 & 29950180 & 29950357 $\pm$ 110 & 29952399 & 29952000 $\pm$ 235 & \\
Pa$^{86+}$ & 31661735 & 31607219 & 31607974 $\pm$ 131 & 31609518 & 31610122 $\pm$ 266 & \\
U$^{87+}$ & 33400172 & 33344927 & 33346253 $\pm$ 169 & 33347312 & 33348148 $\pm$ 220 & \\
Np$^{88+}$ & 35224049 & 35168122 & 35169013 $\pm$ 183 & 35170595 & 35170749 $\pm$ 94 & \\
Pu$^{89+}$ & 37136625 & 37079925 & 37082405 $\pm$ 135 & 37082491 & 37084472 $\pm$ 152 & \\
Am$^{90+}$ & 39143289 & 39085864 & 39087449 $\pm$ 79 & 39088526 & 39090163 $\pm$ 129 & \\
Cm$^{91+}$ & 41247805 & 41189575 & 41192822 $\pm$ 335 & 41192338 & 41194995 $\pm$ 84 & \\
Bk$^{92+}$ & 43455815 & 43396788 & 43400513 $\pm$ 83 & 43399657 & 43403286 $\pm$ 249 & \\
Cf$^{93+}$ & 45771812 & 45711897 & 45715583 $\pm$ 412 & 45714877 & 45718006 $\pm$ 232 & \\
Es$^{94+}$ & 48202067 & 48141250 & 48147673 $\pm$ 39 & 48144347 & 48150157 $\pm$ 168 & \\
Fm$^{95+}$ & 50751624 & 50689785 & 50696869 $\pm$ 408 & 50693004 & 50699333 $\pm$ 89 & \\
Md$^{96+}$ & 53427857 & 53364982 & 53373015 $\pm$ 471 & 53368330 & 53375378 $\pm$ 204 & \\
No$^{97+}$ & 56236972 & 56172980 & 56172549 $\pm$ 383 & 56176463 & 56175545 $\pm$ 145 & \\
Lr$^{98+}$ & 59185705 & 59120454 & 59120944 $\pm$ 44 & 59124080 & 59123626 $\pm$ 337 & \\
\end{tabular}
\end{ruledtabular}
\end{table*}
\newpage

\renewcommand{\thetable}{S2} 
\begin{table*}
\caption{Fine structure splitting of Boron isoelectronic sequence calculated using $H_{\mathrm{DC+B}}$.}
\begin{ruledtabular}
\begin{tabular}{ccccccc}
Atom & $\Delta E_{\mathrm{Ref}}$ & $\Delta E_{\mathrm{CASCI}(10q)}$ & $\Delta E_{\mathrm{BPDE}(10q)}$ & $\Delta E_{\mathrm{CASCI}(18q)}$ & $\Delta E_{\mathrm{BPDE}(18q)}$ & $\Delta E_{\mathrm{Exptl}}$\footnote{NIST Atomic spectra database levels data. \url{https://physics.nist.gov/PhysRefData/ASD/levels_form.html.} Retrieved: 2022-08-22.}   \\
\hline
B & 13.1978 & 11.5561 & 11.7684 $\pm$ 0.7139 & 12.2715 & 12.5287 $\pm$ 0.9813 & 15.287 \\
C$^+$ & 60.457 & 56.316 & 54.401 $\pm$ 2.097 & 58.092 & 61.322 $\pm$ 1.553 & 63.42 \\ 
N$^{2+}$ & 171.73 & 162.92 & 152.35 $\pm$ 4.14 & 166.03 & 166.34 $\pm$ 4.94 & 174.4 \\
O$^{3+}$ & 385.83 & 369.04 & 353.09 $\pm$ 0.24 & 373.80 & 371.88 $\pm$ 3.26 & 385.9 \\
F$^{4+}$ & 750.43 & 720.98 & 695.63 $\pm$ 7.23 & 727.72 & 725.22 $\pm$ 1.89& 744.5 \\
Ne$^{5+}$ & 1322.09 & 1273.76 & 1246.66 $\pm$ 4.82 & 1282.83 & 1223.61 $\pm$ 1.60 & 1306.81 \\
Na$^{6+}$ & 2166.36 & 2091.18 & 2047.99 $\pm$ 3.12 & 2102.97 & 2092.85 $\pm$ 5.20 & 2134.61 \\
Mg$^{7+}$ & 3357.84 & 3245.90 & 3152.15 $\pm$ 3.90 & 3260.83 & 3171.26 $\pm$ 3.70 & 3302 \\
Al$^{8+}$ & 4980.23 & 4819.58 & 4743.64 $\pm$ 59.12 & 4838.05 & 4907.62 $\pm$ 83.88 & 4890 \\
Si$^{9+}$ & 7126.46 & 6902.91 & 6819.95 $\pm$ 49.83 & 6925.39 & 7255.91 $\pm$ 47.77 & 6990.6 \\
P$^{10+}$ & 9898.75  & 9595.83 & 9505.88 $\pm$ 42.88 & 9622.77 & 9835.82 $\pm$ 70.06 & 9699 \\
S$^{11+}$ & 13408.75 & 13007.61 & 13036.97 $\pm$ 14.92 & 13039.49 & 12949.81 $\pm$ 51.71 & 13135.3 \\
Cl$^{12+}$ & 17777.63 & 17257.01 & 17223.49 $\pm$ 61.68 & 17294.32 & 17153.48 $\pm$ 439.42 & 17410 \\
Ar$^{13+}$ & 23136.18 & 22472.44 & 22021.84 $\pm$ 70.65 & 22515.70 & 22210.14 $\pm$ 148.62 & 22656.239 \\
K$^{14+}$ & 29625.02 & 28792.19 & 28760.39 $\pm$ 7.72 & 28841.92 & 28794.81 $\pm$ 38.32 & 29017 \\
Ca$^{15+}$ & 37394.68 & 36364.54 & 36255.11 $\pm$ 16.97 & 36421.27 & 36026.58 $\pm$ 275.92 & 36520 \\
Sc$^{16+}$ & 46605.79 & 45348.01 & 45273.12 $\pm$ 208.58 & 45412.29 & 46394.58 $\pm$ 324.34 & 45637 \\
Ti$^{17+}$ & 57429.26 & 55911.57 & 55957.54 $\pm$ 20.84 & 55983.94 & 56506.46 $\pm$ 410.85 & 56240 \\
V$^{18+}$ & 70046.41 & 68234.83 & 68225.92 $\pm$ 37.58 & 68315.84 & 68274.12 $\pm$ 15.36 & 68610 \\
Cr$^{19+}$ & 84649.24 & 82508.32 & 82487.91 $\pm$ 166.57 & 82598.53 & 82016.14 $\pm$ 389.31 & 82970 \\
Mn$^{20+}$ & 101441 & 98934 & 98606 $\pm$ 28 & 99034 & 97952 $\pm$ 85 & 99360 \\
Fe$^{21+}$ & 120634 & 117724 & 117064 $\pm$ 122 & 117834 & 118606 $\pm$ 71 & 118266 \\
Co$^{22+}$ & 142456 & 139103 & 138991 $\pm$ 61 & 139225 & 139282 $\pm$ 296 & 139290 \\
Ni$^{23+}$ & 167141 & 163309 & 163205 $\pm$ 16 & 163442 & 163367 $\pm$ 69 & 163960 \\
Cu$^{24+}$ & 194940 & 190589 & 190574 $\pm$ 129 & 190734 & 190942 $\pm$ 246 & 191280 \\
Zn$^{25+}$ & 226112 & 221205 & 220977 $\pm$ 5 & 221362 & 220323 $\pm$ 90 & \\
Ga$^{26+}$ & 260931 & 255430 & 254797 $\pm$ 57 & 255600 & 254113 $\pm$ 133 & \\
Ge$^{27+}$ & 299681 & 293551 & 293579 $\pm$ 25 & 293735 & 293665 $\pm$ 75 & 294550 \\
As$^{28+}$ & 342663 & 335869 & 335815 $\pm$ 40 & 336066 & 336175 $\pm$ 190 & 337400 \\
Se$^{29+}$ & 390189 & 382696 & 383035 $\pm$ 66 & 382907 & 383035 $\pm$ 129 & 384460 \\
Br$^{30+}$ & 442583 & 434360 & 434336 $\pm$ 148 & 434587 & 434744 $\pm$ 462 & 436400 \\
Kr$^{31+}$ & 500186 & 491204 & 491490 $\pm$ 44 & 491446 & 491550 $\pm$ 121 & 492560 \\
Rb$^{32+}$ & 563353 & 553584 & 554126 $\pm$ 15 & 553841 & 554563 $\pm$ 297 & 554700 \\
Sr$^{33+}$ & 632452 & 621871 & 621906 $\pm$ 67 & 622144 & 622050 $\pm$ 93 & 623100 \\
Y$^{34+}$ & 707870 & 696452 & 696249 $\pm$ 142 & 696742 & 696679 $\pm$ 152 & \\
Zr$^{35+}$ & 790006 & 777731 & 777783 $\pm$ 149 & 778038 & 778127 $\pm$ 68 & \\
Nb$^{36+}$ & 879278 & 866128 & 865827 $\pm$ 99 & 866452 & 866419 $\pm$ 104 & \\
Mo$^{37+}$ & 976119 & 962077 & 962423 $\pm$ 56 & 962419 & 962080 $\pm$ 70 & 964360 \\
Tc$^{38+}$ & 1080982 & 1066034 & 1065869 $\pm$ 56 & 1066394 & 1066314 $\pm$ 123 & \\
Ru$^{39+}$ & 1194336 & 1178469 & 1179121 $\pm$ 102 & 1178848 & 1179323 $\pm$ 108 & \\
Rh$^{40+}$ & 1316669 & 1299874 & 1300951 $\pm$ 109 & 1300272 & 1301248 $\pm$ 238 & \\
Pd$^{41+}$ & 1448489 & 1430757 & 1430699 $\pm$ 193 & 1431174 & 1431057 $\pm$ 238 & \\
Ag$^{42+}$ & 1590323 & 1571649 & 1571393 $\pm$ 40 & 1572086 & 1571904 $\pm$ 121 & \\
Cd$^{43+}$ & 1742720 & 1723099 & 1723176 $\pm$ 154 & 1723556 & 1723573 $\pm$ 53 & \\
In$^{44+}$ & 1906249 & 1885679 & 1885323 $\pm$ 134 & 1886157 & 1885771 $\pm$ 21 & \\
Sn$^{45+}$ & 2081503 & 2059983 & 2059583 $\pm$ 102 & 2060482 & 2060263 $\pm$ 25 & \\
Sb$^{46+}$ & 2269097 & 2246627 & 2247118 $\pm$ 218 & 2247148 & 2247141 $\pm$ 278 & \\
Te$^{47+}$ & 2469670 & 2446251 & 2446077 $\pm$ 54 & 2446794 & 2446222 $\pm$ 57 & \\
I$^{48+}$ & 2683888 & 2659525 & 2658861 $\pm$ 212 & 2660091 & 2659622 $\pm$ 84 & \\
Xe$^{49+}$ & 2912439 & 2887135 & 2886491 $\pm$ 113 & 2887723 & 2887212 $\pm$ 54 & \\
Cs$^{50+}$ & 3156042 & 3129802 & 3129582 $\pm$ 81 & 3130415 & 3129932 $\pm$ 230 & 3131500 \\
Ba$^{51+}$ & 3415441 & 3388271 & 3387950 $\pm$ 117 & 3388908 & 3388588 $\pm$ 39 & 3390000 \\
La$^{52+}$ & 3691414 & 3663321 & 3663278 $\pm$ 131 & 3663983 & 3663899 $\pm$ 64 & \\
Ce$^{53+}$ & 3984765 & 3955756 & 3955821 $\pm$ 192 & 3956444 & 3956279 $\pm$ 98 & \\
Pr$^{54+}$ & 4296331 & 4266416 & 4266307 $\pm$ 270 & 4267129 & 4267222 $\pm$ 177 & \\
Nd$^{55+}$ & 4626983 & 4596167 & 4595512 $\pm$ 80 & 4596907 & 4596254 $\pm$ 31 & \\
\end{tabular}
\end{ruledtabular}
\end{table*}

\begin{table*}
\begin{ruledtabular}
\begin{tabular}{ccccccc}
Atom & $\Delta E_{\mathrm{Ref}}$ & $\Delta E_{\mathrm{CASCI}(10q)}$ & $\Delta E_{\mathrm{BPDE}(10q)}$ & $\Delta E_{\mathrm{CASCI}(18q)}$ & $\Delta E_{\mathrm{BPDE}(18q)}$ & $\Delta E_{\mathrm{Exptl}}$\footnote{NIST Atomic spectra database levels data. \url{https://physics.nist.gov/PhysRefData/ASD/levels_form.html.} Retrieved: 2022-08-22.}   \\
\hline
Pm$^{56+}$ & 4977628 & 4945922 & 4946187 $\pm$ 80 & 4946690 & 4946621 $\pm$ 317 & \\
Sm$^{57+}$ & 5349202 & 5316613 & 5316863 $\pm$ 373 & 5317410 & 5317433 $\pm$ 123 & \\
Eu$^{58+}$ & 5742695 & 5709234 & 5709494 $\pm$ 100 & 5710060 & 5710148 $\pm$ 174 & \\
Gd$^{59+}$ & 6159116 & 6124790 & 6125085 $\pm$ 238 & 6125647 & 6125513 $\pm$ 52 & \\
Tb$^{60+}$ & 6599537 & 6564358 & 6564535 $\pm$ 36 & 6565246 & 6565470 $\pm$ 183 & \\
Dy$^{61+}$ & 7065057 & 7029033 & 7028710 $\pm$ 55 & 7029953 & 7029302 $\pm$ 128 & \\
Ho$^{62+}$ & 7556830 & 7519972 & 7521113 $\pm$ 240 & 7520925 & 7521955 $\pm$ 128 & \\
Er$^{63+}$ & 8076055 & 8038372 & 8038164 $\pm$ 72 & 8039359 & 8038847 $\pm$ 143 & \\
Tm$^{64+}$ & 8623983 & 8585485 & 8584963 $\pm$ 248 & 8586508 & 8585682 $\pm$ 49 & \\
Yb$^{65+}$ & 9201907 & 9162601 & 9161613 $\pm$ 148 & 9163661 & 9162670 $\pm$ 20 & \\
Lu$^{66+}$ & 9811201 & 9771098 & 9770656 $\pm$ 278 & 9772195 & 9771630 $\pm$ 335 & \\
Hf$^{67+}$ & 10453268 & 10412375 & 10411155 $\pm$ 149 & 10413511 & 10412464 $\pm$ 36 & \\
Ta$^{68+}$ & 11129599 & 11087924 & 11087367 $\pm$ 88 & 11089101 & 11088670 $\pm$ 132 & \\
W$^{69+}$ & 11841730 & 11799282 & 11799251 $\pm$ 162 & 11800500 & 11800183 $\pm$ 44 & 11802000 \\
Re$^{70+}$ & 12591280 & 12548066 & 12547901 $\pm$ 101 & 12549328 & 12548956 $\pm$ 240 & \\
Os$^{71+}$ & 13379921 & 13335944 & 13335230 $\pm$ 210 & 13337251 & 13336476 $\pm$ 286 & \\
Ir$^{72+}$ & 14209438 & 14164708 & 14164993 $\pm$ 58 & 14166062 & 14166361 $\pm$ 304 & \\
Pt$^{73+}$ & 15081655 & 15036176 & 15035743 $\pm$ 98 & 15037578 & 15037037 $\pm$ 200 & \\
Au$^{74+}$ & 15998517 & 15952295 & 15953120 $\pm$ 154 & 15953747 & 15954372 $\pm$ 204 & \\
Hg$^{75+}$ & 16962015 & 16915050 & 16915529 $\pm$ 362 & 16916554 & 16917045 $\pm$ 191 & \\
Tl$^{76+}$ & 17974279 & 17926571 & 17926948 $\pm$ 293 & 17928130 & 17928614 $\pm$ 331 & \\
Pb$^{77+}$ & 19037539 & 18989093 & 18991354 $\pm$ 337 & 18990708 & 18992455 $\pm$ 296 & \\
Bi$^{78+}$ & 20154128 & 20104947 & 20104226 $\pm$ 124 & 20106621 & 20105993 $\pm$ 162 & \\
Po$^{79+}$ & 21326506 & 21276597 & 21275126 $\pm$ 30 & 21278332 & 21276994 $\pm$ 167 & \\
At$^{80+}$ & 22557193 & 22506550 & 22505988 $\pm$ 225 & 22508348 & 22507537 $\pm$ 188 & \\
Rn$^{81+}$ & 23848636 & 23797206 & 23796693 $\pm$ 79 & 23799071 & 23798543 $\pm$ 206 & \\
Fr$^{82+}$ & 25204159 & 25151982 & 25150246 $\pm$ 234 & 25153915 & 25152050 $\pm$ 135 & \\
Ra$^{83+}$ & 26626449 & 26573508 & 26573283 $\pm$ 306 & 26575513 & 26575312 $\pm$ 86 & \\
Ac$^{84+}$ & 28118763 & 28065058 & 28063868 $\pm$ 255 & 28067138 & 28065659 $\pm$ 174 & \\
Th$^{85+}$ & 29684193 & 29629687 & 29629591 $\pm$ 486 & 29631846 & 29632012 $\pm$ 173 & \\
Pa$^{86+}$ & 31326619 & 31271332 & 31270575 $\pm$ 84 & 31273573 & 31272033 $\pm$ 194 & \\
U$^{87+}$ & 33049108 & 32992968 & 32993309 $\pm$ 235 & 32995294 & 32995444 $\pm$ 129 & \\
Np$^{88+}$ & 34856318 & 34799359 & 34799070 $\pm$ 47 & 34801774 & 34801172 $\pm$ 66 & \\
Pu$^{89+}$ & 36751482 & 36693612 & 36694673 $\pm$ 199 & 36696121 & 36697077 $\pm$ 376 & \\
Am$^{90+}$ & 38739941 & 38681195 & 38682438 $\pm$ 173 & 38683802 & 38684875 $\pm$ 340 & \\
Cm$^{91+}$ & 40825427 & 40765721 & 40766742 $\pm$ 492 & 40768431 & 40769554 $\pm$ 410 & \\
Bk$^{92+}$ & 43013532 & 42952865 & 42955744 $\pm$ 173 & 42955682 & 42958575 $\pm$ 432 & \\
Cf$^{93+}$ & 45308712 & 45246987 & 45249757 $\pm$ 288 & 45249918 & 45252590 $\pm$ 142 & \\
Es$^{94+}$ & 47717180 & 47654374 & 47658797 $\pm$ 346 & 47657424 & 47661160 $\pm$ 340 & \\
Fm$^{95+}$ & 50243937 & 50179928 & 50186129 $\pm$ 423 & 50183103 & 50188868 $\pm$ 103 & \\
Md$^{96+}$ & 52896291 & 52831053 & 52837308 $\pm$ 182 & 52834360 & 52841209 $\pm$ 105 & \\
No$^{97+}$ & 55680394 & 55613837 & 55622434 $\pm$ 394 & 55617282 & 55624337 $\pm$ 436 & \\
Lr$^{98+}$ & 58602919 & 58534900 & 58533663 $\pm$ 431 & 58538491 & 58536631 $\pm$ 617 & \\
\end{tabular}
\end{ruledtabular}
\end{table*}

\end{document}